\documentclass[11pt, letter]{article}
\usepackage[margin=1in]{geometry}
\usepackage[utf8]{inputenc}
        
\usepackage[square,numbers]{natbib}
\usepackage{authblk}

\usepackage{complexity}

\usepackage{graphicx}              
\usepackage{amsfonts}              
\usepackage{xurl}
\usepackage{hyperref}
\usepackage{caption}
\usepackage{subcaption}
\usepackage[frozencache,cachedir=.]{minted}
\usepackage{tikz}
\usetikzlibrary{quantikz}
\usepackage{tikz-uml}
\usepackage{amsthm}
\theoremstyle{definition}
\newtheorem{exmp}{Example}[section]

\theoremstyle{plain}

\theoremstyle{definition}

\DeclarePairedDelimiter{\abs}{\lvert}{\rvert}
\let\ket\relax
\DeclarePairedDelimiter{\ket}{\lvert}{\rangle}
\let\bra\relax
\DeclarePairedDelimiter{\bra}{\langle}{\rvert}
\let\braket\relax
\DeclarePairedDelimiterX{\braket}[2]{\langle}{\rangle}{#1 \delimsize\vert #2}
\DeclarePairedDelimiter{\lrang}{\langle}{\rangle}

\newcommand{\Romnum}[1]{\uppercase\expandafter{\romannumeral #1}}

\def\Complex{\mathbb{C}}
\def\Z{\mathbb{Z}}

\begin{document}
\title{On Testing and Debugging Quantum Software}

\author[]{Andriy Miranskyy}
\author[]{Lei Zhang}
\author[]{Javad Doliskani \thanks{This work has been submitted to the IEEE for possible publication. Copyright may be transferred without notice, after which this version may no longer be accessible.}}

\affil[]{Department of Computer Science, Ryerson University \protect\\ Toronto, Canada}
\affil[]{{\{avm, leizhang, javad.doliskani\}@ryerson.ca}}

\date{}

\maketitle

\begin{abstract}
Quantum computers are becoming more mainstream. As more programmers are starting to look at writing quantum programs, they need to test and debug their code. In this paper, we discuss various use-cases for quantum computers, either standalone or as part of a System of Systems. Based on these use-cases, we discuss some testing and debugging tactics that one can leverage to ensure the quality of the quantum software. We also highlight quantum-computer-specific issues and list novel techniques that are needed to address these issues. The practitioners can readily apply some of these tactics to their process of writing quantum programs, while researchers can learn about opportunities for future work.
\end{abstract}

\newpage

\section{Introduction}\label{sec:introduction}
Quantum Computers (QCs) are specialized devices that will be able to solve some problems faster than Classic Computers (CCs)~\cite{bernstein1997quantum, deutsch1985quantum}. This is known as `quantum advantage'. Examples of such problems (originating in various fields of science)  are scalable simulations of quantum systems in physics~\cite{feynman1982simulating}, efficient modelling of chemical reactions~\cite{aspuru2005simulated}, accurate pricing of financial instruments and credit scores~\cite{ORUS2019100028}, and fast breaking of encryption codes in cryptography~\cite{shor1997}.

\subsection{QC timeline} \label{sec:timeline}

The field of quantum computing is young: Feynman introduced the idea of quantum computing in 1982~\cite{feynman1982simulating}; Shor proposed the first practically relevant algorithm (for breaking encryption protocols based on integer factorization) that can be efficiently computed on a QC in 1994~\cite{shor1997}.

It took a while to implement an actual QC. A partnership between academia and IBM created the first working 2-qubit\footnote{
At any point of execution, the state of a CC is given by a vector of bits taking the values of $0$ and $1$. The state of a QC is, however, given by a vector of qubits and bits. Qubit is the basic unit of quantum information, on which QC operate.
We will give a more formal definition of a qubit and compare it with a bit in Section~\ref{sec:quant_comp}.} QC in 1998~\cite{Chuang1998}, but it took the company 18 years to make a 5-qubit QC accessible to the public in 2016~\cite{ibm2016}. 

At present, a few QCs are commercially available. D-Wave started selling adiabatic QC in 2011 (although debate about adiabatic QC being a `true' QC is ongoing\footnote{A hybrid of adiabatic and gate-based QC is promising~\cite{barends_digitized_2016}, but no commercial implementation is available.}~\cite{albash2017}) with the current offerings having $>$~5000 qubits~\cite{DWaveAnn7:online}. 

QCs are also available via fully-managed Cloud services.  IBM gave access to  20- and 50-qubit gate-based superconducting QCs to academic and industrial partners to explore practical applications in 2017~\cite{ibm2017} (and the 65-qubit machine was offered in 2020~\cite{IBMsRoad67:online}). For non-commercial use, IBM offers 5- and 15-qubit QCs via IBM Q Experience online platform based on IBM Cloud (along with local- and Cloud-based simulators)~\cite{ibm_quantum}. Rigetti offered 8-qubit superconducting QC in 2017~\cite{RigettiC96:online}. Google built 72-qubit gate-based superconducting QC in 2018~\cite{google2018}. IonQ introduced ion-trapped 11-qubit QC in 2019~\cite{Wright2019}. Honeywell created ion-trapped 10-qubit QC in 2020~\cite{QuantumC77:online}. Xanadu offered 8- and 12-qubit photonic QCs in 2020~\cite{XanaduRe21:online,CloudPla55:online}. Microsoft provides access to a simulator\footnote{A QC can be simulated on a CC~\cite{ibm_quantum,ms_quantum}. A quantum simulator interprets a mathematical function as part of a physical model~\cite{Johnson2014}; however, it will not yield performance improvement that a QC would provide, as the underlying host system of the simulator is still based on bits rather than qubits (a basic unit of quantum information). Thus, one needs a real QC to reap performance benefits.} of a topological QC via Microsoft Quantum Development Kit~\cite{ms_quantum} (and is planning to give access to an actual QC in the future).

Aggregated Cloud services are starting to appear as well. For example, Amazon Web Services started offering access to QC from various vendors via its Braket service in 2019~\cite{aws_braket_intro}. Currently, it offers D-Wave adiabatic 2048- and 5640-qubit QCs, IonQ trapped-ion-based 11-qubit QC, and Rigetti 32-qubit superconducting QC~\cite{AmazonBr86:online}.

\subsection{QC performance} When discussing the performance of the abovementioned QCs, we have to be mindful of the fact that the performance of the QCs (which are based on different architectures) cannot be compared merely based on the number of qubits that each QC has. Conceptually, it is similar to the fact that we cannot compare the performance of CC based solely on the number of central processing unit (CPU) cores and the cores' frequency.  Standardization of benchmarks for QC is currently in the works by an IEEE Working Group~\cite{ieee_wg}.

One of the measurements of the QC performance, introduced by IBM, is the Quantum Volume~\cite{cross2019}, deemed $V_Q$. This metric combines the number of physical qubits, their inter-connectivity, and measurements error rates. For example, while the Honeywell H1 QC has only 10 ion-trapped physical qubits~\cite{QuantumC77:online}, its $V_Q=128$~\cite{Achievin40:online}. In comparison, the IBM's 27-qubit QC has $V_Q=64$~\cite{jurcevic2020demonstration,IBMDeliv10:online}.

Recently, IonQ introduced a QC with 32 ion-trapped perfect qubits with low gate errors, giving it an expected $V_Q > 4 \times 10^6$~\cite{IonQPres32:online}. They have also introduces another measure of QC performance, called Algorithmic Qubit, defined as `the largest number of effectively perfect qubits you can deploy for a typical quantum program~\cite{ScalingI73:online}.'

\subsection{QC applicability} 
\subsubsection{Theoretical perspective} \label{sec:applicability_theory}
Only those problems falling under the bounded error quantum polynomial time ($\BQP$) class defined in computational complexity theory~\cite{nielsen_chuang_2010} can benefit from the QC architecture. The time complexity of algorithms, which solve $\BQP$ class problems, grows polynomially with the increase of the input size on a QC. On the contrary, the time complexity of the algorithms solving the same problems on a CC is not bounded above by a polynomial function and may grow faster (e.g., exponentially) with the increase of the size of the input. 

Formally, it was shown that the relations between $\BQP$ and other popular complexity classes are as follows: $\P \subseteq \BPP \subseteq \BQP \subseteq \P^{\#\P} \subseteq \PSPACE$, where $\P$ is a polynomial time complexity class, $\BPP$ is a bounded-error probabilistic polynomial time class, $\P^{\#\P}$ is $\P$ with $\#\P$ oracle class ($\#\P$ is a set of counting problems and a class of function problems rather than decision problems), and $\PSPACE$ is a polynomial space class, see~\cite{vazirani2002survey} for details. 

Currently, the consensus (although not formally proven) is that some of the nondeterministic polynomial time ($\NP$) problems do belong to the $\BQP$ set; however, $\BQP$ and $\NP$-complete sets of problems do not overlap (see~\cite{vazirani2002survey,nielsen_chuang_2010} for review). That is, a QC will not be able to solve an $\NP$-complete problem.

\subsubsection{Practical perspective}\label{sec:practical_perspective}

Quantum advantage was demonstrated on a superconducting QC in 2019~\cite{arute2019quantum}. Another demonstration of quantum advantage on a photonic QC was done in 2020~\cite{zhong2020quantum} (although the setup used in the experiment may be difficult to scale up or generalize~\cite{choi2021}).

But when will QCs start solving real-world problems? Quantum chemists are already able to improve simulations of small chemical systems~\cite{kandala2017hardware} and some large ones, albeit with approximations~\cite{tavernelli2020resource}, using the existing QCs. Quantitative finacists will need a machine with $\approx$~7.5K logical qubits to price financial instruments~\cite{chakrabarti2020threshold}. Hackers will need a computer with 20M qubits to break the 2048-bit RSA key in less than a day~\cite{gidney2019factor}. 

Another promising area for application of QCs is machine learning. Modern QCs are already capable of solving simple machine learning problems~\cite{johri2020nearest,broughton2020tensorflow}. They will be able to tackle larger problems as computer size increases. However, we have to be mindful that many subroutines required for machine learning (especially related to the linear algebra computations) will achieve polynomial rather than exponential speedup on QC~\cite{johri2020nearest,DBLP:conf/stoc/Tang19}. Thus, quantum machine learning frameworks (such as Tensorflow Quantum~\cite{TensorFl66:online}) will have to carefully decide\footnote{The same strategy is currently used by CC machine learning frameworks to distribute the workload between CPUs and graphics processing units (GPUs).} which subroutines should be executed on a QC and which should stay on a CC.

To start addressing practical use-cases within the next decade, IBM stated that they `need to at least double the Quantum Volume of our quantum computing systems every year.~\cite{chow2020}' So far, IBM is on track, demonstrating the Quantum Volume of 64 on a QC with 27 qubits in 2020~\cite{IBMsRoad67:online}. By 2023, IBM plans to ship a computer with 1,121 qubits~\cite{IBMsRoad67:online} (with the expectation of proportional Quantum Volume growth).

\subsubsection{Effect of QC on software engineering workloads: vision}\label{sec:vision}

As discussed in Section~\ref{sec:practical_perspective}, currently, the programs for QC are targeting problems from the Science, Technology, Engineering and Mathematics (STEM) domain, e.g., factor integers~\cite{shor1997} or sample boson particles~\cite{giordani_experimental_2018}. They are not of interest to a mainstream consumer. Thus, QCs, at the current stage of their evolution, conceptually resemble computers from the 1940s and 1950s. For example, the Electronic Numerical Integrator and Computer (ENIAC), completed in 1945, was used for comparable tasks: to compute the highest factor of $2^{18}$ and simulate decay of neutron particles during nuclear fusion~\cite{haigh2016eniac}. Peculiarly, this parallel is further supported by the fact that languages designed for QC operate at the level of qubits and quantum circuits~\cite{heim2020quantum}.

Does this mean that history will repeat itself, and a new Software Crisis~\cite{DBLP:conf/qce/MoguelBGM20}, similar to the one that led to the inception of Software Engineering in the 1960s, is upon us? The authors' position is cautiously optimistic: below, we argue that this is not the case and that the history of computing evolves in an upward spiral rather than a circle.

On the surface, the current situation with programming QC is similar to programming CC in the 1950s: we are dealing with expensive machines which have to be manipulated at the level of registers and gates. A highly-qualified personnel is required for the machine's maintenance. However, the situation is much better than back in the day. We now understand how to deal with large codebases, e.g., by applying Lehman's laws of software evolution~\cite{lehman1980}. The tools and access to resources have also improved dramatically. Punched cards and fights for machine time have been replaced with powerful integrated development environment (IDEs) and trivial access to computing resources (as a lot of QC development can be done in a simulator). Thus, programming modern QC is a much more pleasant and forgiving experience than that of early generations of CCs.

For example, we are programming QCs at the gate level, but we are typically doing from higher-level languages. For example, the QisKit library is coded in Python. Thus, we get superior language constructs, such as for-loops, modularity, or classes\footnote{A reader interested in exploring high-level language constructs may want to examine the source code~\cite{qiskit:shor:online} of the \texttt{Shor} class, used in Figure~\ref{fig:aqua_shor}.}. We can build automatic test cases using test harnesses designed for these high-level languages. We use potent IDEs with integrated source code management, code-completion, and spell-checking. The code can be executed in the simulators and, in some cases, debugged using an interactive debugger (e.g.,~\cite{ibm_quantum,ms_ide}). Finally, we have amassed decades of knowledge on requirements engineering and design --- a lot of this knowledge can and will be transferable to the QC programming (see~\cite{DBLP:journals/corr/abs-2007-07047} for review of the latest works). We will also show how we can readily transfer some of software engineering (SE) knowledge, especially in the context of System of Systems (SoS), in Sections~\ref{sec:usage} and \ref{sec:qa}.

Ironically, the ability of future QCs to break modern encryption schemes may trigger a crisis related to the maintenance of the legacy software executed on a CC (rather than a crisis of developing code for QCs). While the breakage of RSA keys discussed in Section~\ref{sec:practical_perspective} is years away, we need to start protecting ourselves against these future potential attacks right now. This is because malicious entities can harvest sensitive data communications now and~--- when a powerful QC becomes available~--- leverage that computing power to break today’s non-quantum-resistant encryption and gain access to such sensitive data. Many QC-resistant encryption methods have been proposed (see~\cite{zhang2020quantum} for details). However, their implementations will require significant changes to various software, such as web browsers and web servers, mail and hard drive encryptors. We conjecture that the amount of work required to introduce these changes into legacy software may be similar to that of the Y2K problem~\cite{britannica_y2k}. For details on protection steps, see~\cite{zhang2020quantum}.

\subsection{Overview of the rest of the paper}\label{sec:overview}

The programming languages for QC are mainly low-level, operating at the level of QC register, e.g., Open Quantum Assembly (OpenQASM) language~\cite{cross2017open}. However, higher-level languages are being developed (e.g., Scaffold~\cite{JavadiAbhari2014ScaffCC}). 

To enable usage of the QC, libraries with pre-packaged quantum algorithms start to appear. For example, Qiskit Aqua~\cite{Qiskit} (an open-source library written in Python) implements quantum algorithms for various domains, such as artificial intelligence, chemistry, and finance. Such a library enables a programmer to treat QC as a black-box and leverage quantum algorithms without having a deep understanding of the QC field. We will discuss the usage of quantum components based on such libraries as part of a software solution in Section~\ref{sec:usage}. We will then cover the implications of quantum components to testing and debugging SoS and standalone QC programs in Section~\ref{sec:qa}. 

The quantum libraries themselves have to be developed by programmers with expertise in the QC field. These programmers, inevitably, inject defects in their code (uniting CC and QC programming worlds). After that, the code has to be debugged. We will touch on existing debugging tactics and their applicability to quantum programs in Section~\ref{sec:traditional}.  To better understand programmers' challenges, we will review and compare classic and quantum models of computation in Section~\ref{sec:quant_comp}. Armed with this knowledge, we will then show tricks for analyzing quantum  programs during runtime in Section~\ref{sec:debug_quantum}. We conclude the paper in Section~\ref{sec:conclusions}.

\section{Creation and Usage of Quantum Software Components}\label{sec:usage}

As discussed in Section~\ref{sec:applicability_theory}, $\P \subseteq \BQP$. Thus, one may argue that QCs will replace CCs at some point in time. However, we conjecture~\cite{miranskyy2019testing} that QCs will not replace CCs in the short run. Rather, QCs will be integrated into an SoS, where QC-based components will solve $\BQP$ problems (that CCs cannot solve), while the solution will be passed to CC components for post-processing. Let us elaborate on this conjecture.

The reasons for this lie in economics. Modern QCs are expensive: e.g., D-Wave QC is valued at \$15 million~\cite{wired2017}. Many require low-temperature cooling and specialized training to use and maintain. The QCs are bulky, taking significant amount of space. The costs, size, and maintenance requirements will probably be reduced over time (as it happened during the transition of mainframes to personal computers). Let us speculate how various QC architectures may evolve in the future.

The superconducting QCs, such as the ones from IBM, Google, and Rigetti, require cryogenic cooling. This is the fundamental physics requirement and will not change with time. Thus, even though such QCs may become smaller and cheaper, they will require specialized cooling and maintenance personnel. Thus, these machines will have to be hosted in a public or private Cloud and accessed as a service.

There are at least two QC architectures for Quantum Processing Units (QPU) that may one day be integrated into a CC, similar to a GPU. The first one is ion-trap QC, such as the one from Honeywell and IonQ, which do not require cryogenic cooling (ion trap QC use vacuum and lasers to `slow down' atoms). IonQ plans to have a rack-based QC by 2023 and a desktop unit by 2025~\cite{IonQplan81:online}.

The second architecture is photonic, such as the one from Xanadu. The photonic QC chip operates at room temperature~\cite{arrazola2021quantum}. However, photon detectors in photonic QC do require cooling~\cite{choi2021}. Potentially, engineers may be able to come up with a sensor that does not require cooling. This will allow photonic QC to be miniaturized and operate at room temperature.

Another argument for using QC in an SoS comes from the cost of software development. Theoretically, one can port any CC code to a QC code. However, the cost of porting will make it economically infeasible. Modern QC programming languages, such as IBM QisKit Python package~\cite{ibm_quantum}, Google Cirq framework~\cite{cirq:developers:2021:4586899}, or Microsoft Q\# language~\cite{svore2018}, operate at the level of qubits and quantum circuits. Creation of, e.g., a graphical user interface, in such a language would be very time-consuming and expensive\footnote{Notwithstanding, these languages integrate nicely into CC domain, simplifying the creation of SoS. As mentioned above, QisKit is implemented as a Python library, running on a CC. Once translated to QC machine language (via OpenQASM), the code is passed to the QC for execution (the complexities of the call are encapsulated in the library’s code). Cirq and Q\# behaviour is similar: the code is developed on a CC and then passed to a QC for execution.}. 
In the distant future, as the higher-level languages for QC are created, the replacement of CC with QC will become more probable. 

For now, it is easier to keep the existing code on CC and outsource parts that can run efficiently on a quantum machine to a QC. How can this be done? Let us look at an example.

\begin{exmp}\label{ex:sos}
Suppose that we need to create a software-as-a-service for factoring large integers to break the RSA cryptosystem. The time complexity of the best algorithms available for a CC in the family of general number field sieves) is sub-exponential~\cite{pomerance96atale}. Thus, these algorithms will be ineffective for large integers. Instead, we will build a software component running Shor’s algorithm on a QC, which will be more efficient for large integers, because Shor’s algorithm computation time (as other $\BQP$ class algorithms) will grow polynomially with the growth of the input integer $N$ (when executed on a QC). The rest of the components, such as user interface (UI) and application program interface (API) for obtaining input (i.e., the value of $N$) to be passed to the QC component and to return the vector of factors $\vec{L}$ back to the user will be implemented on a CC, as depicted in Figure~\ref{fig:arch}.

Modern libraries that abstract QC computations can already enable this scheme. For example, the QisKit Aqua library (written in Python) already has Shor's algorithm built-in~\cite{qiskitaq4:online}. Thus, a programmer does not need to know anything about quantum algorithms and the implementation details. Instead, they will simply call the Python class implementing Shor's algorithm. We show a sample implementation of this approach in Figure~\ref{fig:aqua_shor}.

\begin{figure*}[tb]
    \centering
    \resizebox{0.9\linewidth}{!}{
    \begin{tikzpicture}[outer/.style={draw=gray,dashed,fill=green!1,thick,inner sep=5pt}]
    \begin{umlseqdiag}
    \umlactor[scale = 0.5]{user}
    \umlbasicobject{WebApp's UI}
    \umlbasicobject{WebApp's Backend}
    \umlbasicobject{QC's Controller}
    \umlbasicobject{QC's Core}
    \begin{umlcall}[op=$N$, type=synchron, return=Return $\vec{L}$]{user}{WebApp's UI}
    \begin{umlcall}[op=$N$, type=synchron, return=Return $\vec{L}$]{WebApp's UI}{WebApp's Backend}
    \begin{umlcall}[op=QASM, type=synchron, return=Return $\vec{L}$]{WebApp's Backend}{QC's Controller}
    \begin{umlcall}[op=Control sequence, type=synchron, return= Return qubits' state]{QC's Controller}{QC's Core}
    \end{umlcall}
    \end{umlcall}
    \end{umlcall}
    \end{umlcall}
    \node (text1) [anchor=north] at ([xshift=5.5em, yshift=1.5em]WebApp's UI.north) {WebApp};
    \node (text2) [anchor=north] at ([xshift=5.5em, yshift=1.5em]QC's Controller.north) {QC};
    \begin{pgfonlayer}{background}
    \node[outer,fit=(WebApp's UI) (WebApp's Backend) (text1)] (A) {};
    \node[outer,fit=(QC's Controller) (QC's Core) (text2)] (B) {};
    \end{pgfonlayer}
    \end{umlseqdiag}
    \end{tikzpicture}
    }
    \caption{Sequence diagram for Example~\ref{ex:sos}. A user submits the value of integer $N$ for factorization via UI of a Web App, which passes $N$ to the WebApp's backend. At the backend where the value of $N$ is passed to, Shor’s algorithm is implemented using, say QisKit library~\cite{ibm_quantum}. The library translates QisKit code into OpenQASM and passes QASM listing to the Controller of a QC. The Controller, which initializes the QC Core based on the OpenQASM code, triggers its execution and measures the values of the qubits once execution ends. The Controller converts the measurements into the elements of $\vec{L}$. These values are then returned to the Backend, UI, and, finally, the user. The sequence is depicted as synchronous, but can be made asynchronous if required by a use case. Note that the WebApp UI and Backend, as well as the QC Controller, are running on CCs. The QC Core represents the `true' QC. However, the QC Controller and the QC Core can be thought of as one QC system from practical perspective. The figure and example are adopted from~\cite{miranskyy2019testing}.}
    \label{fig:arch}
\end{figure*}
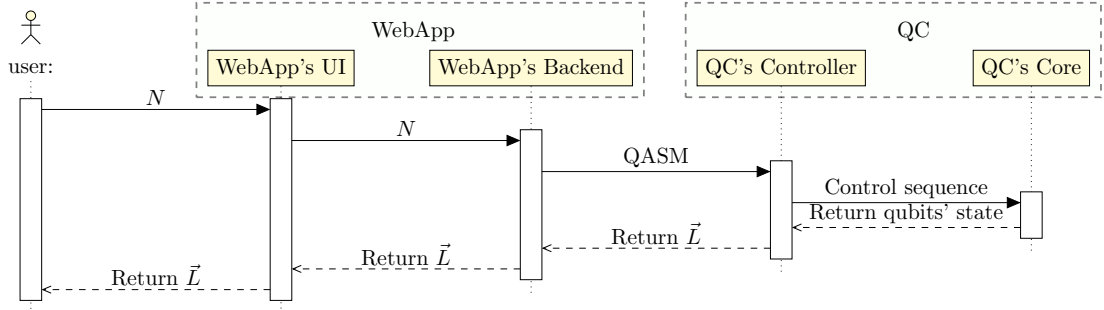

\begin{figure}[tbh]
\centering
\begin{minted}[xleftmargin = 4mm, fontsize = \scriptsize, numbersep = 2mm, linenos = true]{python3}
from qiskit import Aer
from qiskit.aqua.algorithms import Shor

def factorize_integer(my_int):
    # Specify the backend/QC,
    # which will be used for computations.
    # here, we pick a simulator rather than a live QC.
    backend = Aer.get_backend('qasm_simulator')
    # Factor the integer N, 
    # a is a random integer that satisfies 
    # a < N and gcd(a, N) = 1.
    algorithm = Shor(N = my_int, a = 2)
    result = algorithm.run(backend)
    return result['factors']

int_to_factor = 15
print(f"The factors for {int_to_factor} are " 
      f"{factorize_integer(int_to_factor)}")
# Output:
# The factors for 15 are [[3, 5]]
\end{minted}

\caption{Factor integer $N=15$ using Qiskit Aqua Python package. The code can be implemented as a one-liner, but we split it into multiple code lines to improve code comprehension.}\label{fig:aqua_shor}
\end{figure}

\end{exmp}

The above example can be easily integrated into any existing SoS. For example, a microservice on a CC can be instantiated to make an asynchronous call to a QC backend offered as a managed Cloud service (such as the ones discussed in Section~\ref{sec:timeline}). 

The same approach will be readily applicable to rack-based or desktop-based QPU units when they become available. In this case, the backend will simply point to a local rather than a remote device.

\section{Quality Assurance}\label{sec:qa}
In this section, we will look at various viewpoints on quality assurance. We start with a comparison of the white- and black-box testing in Section~\ref{sec:wb_bb}. Then, we explore black-box testing focusing on using QC as a component in an SoS in Section~\ref{sec:bb_sos}, followed by the black-box testing of the QC component itself, concentrating on verification and validation in Section~\ref{sec:vv}. Finally, in Section~\ref{sec:mapping}, we discuss mapping to test phases of the activities covered in Sections~\ref{sec:wb_bb}--\ref{sec:vv}.

\subsection{White- and black-box testing}\label{sec:wb_bb}

Two widespread methods of testing are white- and black-box testing. The former method tests internal data structures and program flow. The latter method tests the functionality, ignoring the inner workings of the software, answering the following question: will we get an expected output for a given input?

We can perform all of the standard white-box activities on the code listing, such as code reviews and code inspections. We can build linting and code inspection tools similar to those available for CC languages and run them automatically in an IDE.

However, interactive debugging (another popular white-box activity) is challenging by construction because a QC is a black-box. Based on the classical quantum mechanics, we cannot observe the inner working of a program (executed on a QC) without altering the program’s state and the final result, as measuring a qubit destroys superposition~\cite{kaye2007introduction}. 

This implies that, currently, we cannot perform interactive debugging of a program running on a QC, as we have to stop the program and take the measurements. Having said that, we may be able to debug the code in a CC simulator\footnote{For example, Microsoft Q\#~\cite{svore2018} provides language constructs to define facts and assertions, and take registry measurements that can be visualized in the Microsoft Visual Studio IDE~\cite{ms_ide}.  } if the CC is powerful enough to perform the computations. 

Moreover, suppose the QC program can be decomposed into modules, and a given module produces the measurable expected output. In that case, we can write xUnit test cases for this module that can be tested in a simulator~\cite{svore2018} or on a real QC\footnote{Given that most of QC architectures are noisy, we have to run the program multiple times and compute an expected value and the confidence interval for these measurements, which may require the creation of probabilistic test cases similar to~\cite{DBLP:conf/icse/Honarvar0N20}. This comes at a cost~\cite{miranskyy2019testing}, see Section~\ref{sec:verification} for details.}. 
We can even estimate the coverage that our test cases provide using input and output coverage criteria~\cite{ali2021assessing}. Finally, we can apply clever tricks, such as separate qubits into subgroups and measure them individually to perform approximate measurements; we will discuss these tricks further in Section~\ref{sec:debug_quantum}.

Yet, it is often easier to resort to black-box testing when dealing with a program running on an actual QC. Let us explore the black-box testing further.

\subsection{Black-box testing: component in an SoS}\label{sec:bb_sos}

As we discussed in~\cite{miranskyy2019testing} and reiterated in Section~\ref{sec:usage}, we believe that for the foreseeable future, the QC will be used as a component in an SoS, where the code running on a CC will request to compute parts of the code on a QC component. Using a REST interface, we pass the request to the QC (residing in a public or private Cloud) and then get the response to our request via the same interface. The interface is abstracted and hidden in a library, e.g., QisKit, so that the programmer does not have to worry about the details of the interface.

Let us revisit Example~\ref{ex:sos}, which depicts this flow in a UML sequence diagram in Figure~\ref{fig:arch}, where the app programmer’s `visibility' stops at the WebApp’s Backend level. An example of such backend code is given in Figure~\ref{fig:aqua_shor}. The programmer interacts with the QC as with any generic Platform as a Service (PaaS) via an API (hidden in the QisKit library).

If we operate at this `granularity level', then we have the full power of existing SE tools at our disposal. For example, we can build UML diagrams to understand the relationship between components, interaction sequences, or system states. Note that we do not need to extend the UML notation at this level of granularity\footnote{For the actual quantum code, UML extensions might be useful~\cite{DBLP:conf/icse/Perez-DelgadoP20}.}: an architect/designer with no specialized training in quantum computing can create such diagrams. These diagrams, along with the specifications and requirements for the SoS, can help a tester create test plans without any knowledge of the QC.

When testers create automatic test cases, they can either use a simulator backend or, which may be even more efficient, use test doubles and replace calls to the QC with mocks or stubs~\cite{meszaros2007xunit}. For example, going back to the code in Figure~\ref{fig:aqua_shor}, as QisKit is written in Python, we can use our favourite Python Mock library (e.g., a built-in library unittest.mock~\cite{unittest43}) to create test-doubles for the calls to the QC on lines 8, 12, and 13. Alternatively, we can generate a test-double for the whole function \texttt{factorize\_integer}.

The difficulty of creating the unit test may depend on the type of output returned by the QC. Let us explore this further.

\subsection{Black-box testing: verification and validation}\label{sec:vv}

When testing the programs, how can we ensure that our code follows the design document and that the QC is doing what it is supposed to do? And even if our code reflects the design, how can we make certain that the output of the program delivers what a user needs? The former will be discussed in Section~\ref{sec:verification}, the latter --- in Section~\ref{sec:validation}. We will conclude with examples of code validation in Sections~\ref{sec:val_p} and~\ref{sec:val_super-p}.

\subsubsection{Verification}\label{sec:verification}

As discussed in Section~\ref{sec:wb_bb}, we can apply the full spectra of verification techniques on the code listings, but verification of a running program on an actual QC is more challenging, as we cannot properly use interactive debugger. To verify the correctness, we can try to run and debug our program in a local or online simulator, such as~\cite{ibm_quantum,ms_quantum}. However, as the simulators run on CC, we will have to limit the complexity of our test cases to obtain results within a reasonable amount of time. This will help us to eliminate some of the defects (a taxonomy of QC bugs is being developed~\cite{huang2018qdb}), but does not guarantee that no other defects will be encountered while solving production-scale problems. The same issue, conceptually, arises with CCs too, e.g., when dealing with buffer-overrun- and resource-leak-related defects.

The above test assumes that a simulator correctly and accurately resembles the actual QC, which is not always the case. Thus, a more definitive verification of correctness should be done on the the actual QC. Given the probabilistic nature of QC, we may have to execute the same code multiple times to increase the accuracy of our answer using Chernoff bound~\cite{nielsen_chuang_2010} (which is similar to probabilistic algorithms in $\BPP$ class running on CC~\cite{nielsen_chuang_2010}). This repeating functionality is built into packages like QisKit, but it requires a higher amount of computing resources (proportional to the number of repetitions).

The above approach assumes that the QC hardware, its operating system, and the compiler/translator of our program are running correctly, which is not always the case. To verify their correctness, we may need to execute the same program on multiple QCs (preferably from different manufacturers) and compare the results. If results differ significantly --- it may be a sign that there is an issue with one or more QCs under test. This is akin to correctness testing of a database engine by running the same query against different database engines~\cite{cialini2007method}.

An award-winning protocol, verifying QC computations with the help of a CC has been proposed~\cite{mahadev_2018}. It requires a significant amount of computational resources and, probably, will not be implemented shortly. However, as the computing power of QCs will increase, this protocol will become implementable in practice.

Finally, even if all of the above tests pass, it does not guarantee that the actual results returned by the QC are correct. This is where validation comes into play.

\subsubsection{Validation}\label{sec:validation}

When doing the validation, we need to make sure that the output of the algorithm satisfies the conditions provided in the requirements document (assuming that requirements were captured correctly). In other words, validation of a program running on a QC is similar to that of a program executed by a CC. Essentially, the ease of validation will depend on the difficulty of implementing a program for validating the results and the time needed\footnote{As discussed in Section~\ref{sec:applicability_theory}, many problems in $\BQP$ are solved efficiently on a QC, but are challenging to solve on a CC. However, the time needed to solve a problem is independent of the time needed to validate this solution.} to execute the validation. 

Before implementing a program, we need to estimate how long the validation process would take. To do so, we can resort to complexity analysis. Say, if the execution time of the validation\footnote{In the algorithm-related literature, the term verification rather than validation is used. We will use the term validation for consistency with the name of this section.} program would belong to $O(1)$, the validation process (given that it is easy to code up) would be straightforward. However, if the execution time would belong, say, to $O(n!)$, where $n$ would be proportionate to the length of input into the validation process (and to the length of the solution), then the validation process for a significantly large $n$ would be formidable. 

For simplicity, we can split the complexity of validation into two classes: polynomial time $\P$ bounded by $O(n^k)$ (i.e., validation time is bounded above asymptotically by $n^k$, where $k>0$) and super-polynomial time $\P^\C$, which is complementary to $\P$, bounded by $\omega(n^k)$ (i.e., validation time dominates asymptotically the $n^k$). We will look at examples of the algorithms belonging to these classes in subsections below.

Where should we implement the validation program: on a CC or a QC? In the program belongs to  $\P$ class, it can be implemented on either one, as $\P \subseteq \BQP$. However, as discussed in Section~\ref{sec:usage}, it is challenging to program a QC as we are dealing with low-level programming language. Moreover, the cost of running a QC in comparison with a CC is high. Thus, it is simpler and more economically feasible to implement a validation program on a CC, if possible.

In the case of $\P^\C$ class, the answer is less straightforward. If the size of the input into validation program is small, we may be able to still leverage a CC (especially if we can parallelize the validation on a CC cluster). However, we may have to resort to a QC for larger problems. If the validation program belongs to classes which are a subset of $\BQP$, such as $\BPP$ class, then QC is a good match. However, if the validation belongs to a `harder' class, such as $\NP$-complete, then QC may also fail to deliver timely results. In this case, we may have to resort to a heuristic that tries to roughly validate the solution, but does not guarantee that solution is correct.

Let us look at examples of algorithms from both classes and ways to run the validation.

\subsubsection{Validation: Polynomial $\left( \P \right)$}\label{sec:val_p}

\begin{exmp}\label{ex:shor}
Shor’s integer factoring algorithm (which we used in Example~\ref{ex:sos}) takes integer $N$ as input and returns a vector of prime factors $\vec{L}$ for $N$~\cite{shor1997}. The solution runs on a QC in polynomial time, $O((\log N)^2 (\log \log N) (\log\log\log N) )$ to be specific~\cite{shor1997}. The validation complexity is independent of the solution complexity,  growing linearly with the number of elements in $\vec{L}$, deemed $l$, as we simply need to multiply the elements in $\vec{L}$ to do the validation. That is, the complexity of validation of Shor’s algorithm is $O(l)$. Thus, we can easily\footnote{Although we may have to leverage existing libraries for multiplication of integers with arbitrary precision, such as Java's BigInteger~\cite{BigInteg48}.} perform validation on a CC. 
\end{exmp}

\begin{exmp}
Grover’s algorithm~\cite{grover1996fast} takes an unsorted list of $M$ items, out of which there is one item with a unique property (e.g., a unique value) that we would like to retrieve. The algorithm returns the index $i$ of this item of interest. Its  complexity on a QC is $O(\sqrt{M})$, while complexity of the fastest solution running on a CC is $O(M)$~\cite{grover1996fast}. Suppose that the item is retrieved at $i$; then the time complexity of validation of Grover’s algorithm is $O(1)$, because we only need to perform one evaluation of the item at $i$. This can be easily done on a CC. 
\end{exmp}

Technically, validation of the two algorithms can all be carried out on a QC. But this is economically inferior, as discussed above.

\subsubsection{Validation: Super-polynomial $\left( \P^\C \right)$}\label{sec:val_super-p}

\begin{exmp}
Boson sampling is a good example of a problem that is challenging to validate. Yet, the algorithm is crucial\footnote{It may lead to the implementation of a non-universal QC, which will still be more efficient than CC for some tasks, see~\cite{aaronson2011} for details.}. Experimentally, the algorithm is typically implemented using photons (belonging to the family of boson particles~\cite{nielsen_chuang_2010}). To implement the algorithm, we need a linear-optical circuit with $m$ modes that is injected with $h$ individual photons ($m>h$)~\cite{giordani_experimental_2018}. In this implementation, the boson sampling task reduces to creating a sample from the probability distribution of individual photon measurements at the circuit’s output. 

This algorithm cannot be computed on a CC for large values of $m$ and $h$, as it requires computing a permanent of a matrix which is a $\#\P$-hard problem~\cite{aaronson2011,valiant1979}. At best, it requires $O(h 2^h + mh^2)$ operations~\cite{clifford2018}.

However, the problem does fall~\cite{aaronson2011} into $\PostBQP$ class ($\BQP$ class with post-selection), which can be efficiently computed on a QC. Validation of the results on a CC is also a $\#\P$-hard problem, as we again need to compute the permanent of a matrix. However, one may adopt a heuristic to estimate goodness of findings (essentially, performing approximate validation) using machine learning approach~\cite{giordani_experimental_2018}.
\end{exmp}

\begin{exmp}\label{ex:gauss}
The second example is estimating Gauss sums. The classical algorithm for estimating Gauss sums on CCs (for polynomials of degree $\geq 3$) belong to the  $\#\P$-hard class~\cite{cai10}. However, QCs can estimate Gauss sums to polynomial precision in polynomial time~\cite{van2002efficient}. To validate the Gauss sums on a CC, we need to take the same inputs and compute the Gauss sum, which, as we know, is a  $\#\P$-hard problem. Hypothetically, one may invent an approximate validation heuristic (like in the boson sampling case discussed above); to the best of our knowledge, none exist at the time of writing.   
\end{exmp}

In the above examples, to perform an accurate validation, we need to do it on a QC. Ideally, this should be done on a different QC to simultaneously check the correctness of the computer itself (as was discussed in Section~\ref{sec:verification}). The code of the validation software would be similar to the one of the solution software. Thus, if resources permit, one may want to create the validation code from scratch (rather than reusing the existing code from the solution) to avoid migration of the defects from the solution code into the validation code.

\subsection{Mapping of activities to test phases}\label{sec:mapping}

Let us now discuss which testing phases we can apply to a QC. Techniques for some of the later test phases are readily transferable from the CC domain. For example, the performance quality assurance team can time the execution of the software on a QC to detect performance degradation during the performance testing phase. Another example is users helping to uncover defects by reporting failures during one of the acceptance test phases (say, beta testing).

If we treat a QC as a black-box component in an SoS, based on the discussion in Section~\ref{sec:bb_sos}, the situation is straightforward. We can perform integration testing of a QC component and then see how it behaves in the SoS during the execution of a test scenario during system testing. Given that the QC component is black-box, the testing will be no different from the testing incorporation of yet another PaaS component into the SoS.

However, dealing with testing the QC component itself is more convoluted. Let us explore three core testing phases: unit testing (UT), functional testing (FT), and system testing (ST). 

Unit tests in the CC domain can leverage black- and white-box testing.  Here, we can use the techniques discussed in Section~\ref{sec:wb_bb}. For the white-box testing, we may be able to debug parts of the QC program on a simulator or a QC. For the black-box (or grey-box) unit testing, we can create test cases for the modules that return a measurable value. 

What about FT? As we discussed in Section~\ref{sec:vision}, QC machines are targeting STEM-oriented problems with relatively simple inputs and outputs (for concrete cases revisit Examples~\ref{ex:shor}--\ref{ex:gauss}). 
Thus, the problems, which algorithms on QC are trying to solve, are self-contained. In other words, an algorithm implemented on QC represents a piece of functionality. Unit testing typically focuses on individual units/modules of code, where a combination of these modules yields a piece of functionality. FT, on the other hand, tests a particular piece of functionality. This implies that when we perform black-box testing (i.e., checking the correctness of output for a given input) on a program that implemented such an algorithm on a QC, we are doing FT rather than UT. That is, if we are to test expected output of function \texttt{factorize\_integer} in Figure~\ref{fig:aqua_shor}, we may classify this test as FT rather than UT.

What about ST? There exists a number of (sometimes conflicting) definitions of FT and ST. Let us adopt the definition that FT `verifies a program by checking it against ... design document(s) or specification(s)' while ST `validate[s] a program by checking it against the published user or system requirements'~\cite[p. 52]{kaner1999testing}. In the case of usage QC component in SoS, QC output can be tested in isolation to check it against design and specs, hence the FT. Then we can perform inegration testing and then execute QC component in a test workload as part of ST efforts.

What if running QC component standalone (akin to Figure~\ref{fig:aqua_shor}) is all that is required? This is not uncommon for STEM use-cases, where we often would like to compute and save a value. In that case, the boundary between FT and ST blurs. We may argue that we are doing ST (of a small system) rather than FT, as we are mainly checking the correctness based on requirements. For example, if we are to implement Shor's integer factorization algorithm on a QC, the requirement would be as follows. `Given a composite number $N$ as input, the algorithm should output a vector of integers $\vec{L}$ (with the integers strictly between $1$ and $N$), such that the product of these integers is equal to $N$.' To check the correctness of our implementation of Shor's algorithm, we will multiply the returned integers and compare the resulting product against $N$, thus validating our implementation of the algorithm. Note that we resort to a requirements document rather than a design document to check correctness, thus performing the ST rather than the FT. We can also perform verification, hence the FT, by checking the reproducibility of the results.

\section{Debugging Tactics}\label{sec:traditional}
Debugging is a process of removing an error, once this error has been exposed and is often a consequence of successful testing practices~\cite{pressman2014software} (some of which we discussed above). While we hope that one day debugging will become an orderly and automated process (e.g., by automatically mapping bug reports to code where the defect resides, and then issuing a patch for this code~\cite{TufanoPWBP19}), currently it is an art more than a science~\cite{pressman2014software}. 

The high-level tactics~\cite[Chapter 8]{myers2011art} for debugging a software had not changed significantly over the last 42 years (when the first edition of the seminal work~\cite{myers2011art} was published), although integrated development environments and various automation tools have streamlined a lot of mundane tasks~\cite{zeller09debugging, MargineanBCH0MM19}. The three common tactics~\cite{myers2011art,pressman2014software} are backtracking, cause elimination, and brute force, discussed below.

\textit{Backtracking} debugging centers around examining the execution tree from the point of the error until a perpetrating code block is found. The analysis techniques for a code listing (such as code reviews and inspections) of a CC program can be readily applied to a QC program~\cite{miranskyy2019testing}. Thus, these tactics are transferable. Anecdotally, based on discussions with practitioners, code reviews and inspections are the most popular debugging techniques of quantum programs nowadays.

\textit{Cause elimination} debugging formulates a hypothesis (using inductive or deductive reasoning), specifying a root cause for a bug under study. Then, data are devised, and experiments are conducted to refute or prove this hypothesis. This approach can be applied to QC. Given the probabilistic nature of the QC programs~\cite{nielsen_chuang_2010, miranskyy2019testing}, we will have to execute the program multiple times to obtain a distribution of the results and assess the accuracy of the answer. Thus, we may be able to extend the techniques used for testing probabilistic programs running on CC, such as~\cite{DuttaLHM18, DuttaZHM19}, to the QC domain. Such techniques already start to appear~\cite{huang2019statistical,DBLP:journals/pacmpl/LiZYDY020,DBLP:conf/icse/Honarvar0N20}.

\textit{Brute force} debugging --- centred around the analysis of runtime traces, memory dumps, and output statements ---  focuses on runtime data analysis. Of the three tactics, this is the most common one~\cite{pressman2014software}. Some of the analyses of the runtime artifacts can be automated; however, a lot of the brute force debugging is still performed manually~\cite{pressman2014software}. Can we transfer these tactics? 

If we treat a QC program as a black-box, then the short answer is `yes'.  As discussed in Sections~\ref{sec:usage} and \ref{sec:bb_sos}, if a QC program will be used as part of an SoS, then we can trace the input (passed from the CC component to the QC component) and the output (from the QC component to the CC component). The input and output data can be recorded in a log, and these data can be compared against the expected values. 

But what if we would like to analyze a QC program at runtime using a white-box approach, e.g., to capture the execution trace of a QC program or perform interactive debugging of the code executed on the QC? In such a case, the short answer is `it depends'~\cite{DBLP:conf/icse/Miranskyy0D20}. Before we delve into the answers, let us compare and formally define classic and quantum models of computation, which will help us understand the issues with debugging quantum programs.

\section{Quantum Computation}\label{sec:quant_comp}

In this section, we review the basic concepts of quantum computation and set up the conventions and notions that are used in the rest of the paper.

\subsection{The classical model of computation}
\label{sec:cls-cmm-model}

\begin{figure}[ht]
    \centering
    \begin{quantikz}[row sep = {0.8cm,between origins}]
        \lstick{$x_0$} & \ctrl{1} & \gate{G_2} & \gate[wires = 2]{G_3} & \qw & \qw \\
        \lstick{$x_1$} & \gate[wires = 2]{G_1} & \qw & & \gate{G_5} & \qw \\
        \lstick{$x_2$} & & \qw & \gate{G_4} & \octrl{-1}
    \end{quantikz}
    \caption{A classical circuit.}
    \label{fig:cls-crct}
\end{figure}
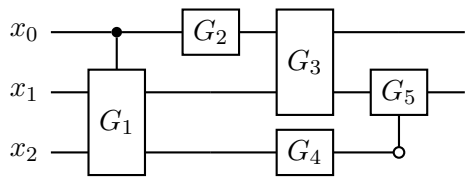

Classical computation can be modeled using the language of circuits. Abstractly, a circuit is a network of gates and wires: the wires transmit bits to gates, the gates perform elementary operations on the input bits, and the results of these operations are again bits that are carried by wires. Figure \ref{fig:cls-crct} shows a schematic example of a classical circuit. 

An important concept regarding the circuit model of computation is \textit{universality}. A set $G$ of gate are said to be universal if given any function $f: \{0, 1\}^m \rightarrow \{0, 1\}^n$, where $m$ and $n$ are positive integers, a circuit for $f$ can be constructed using only gates from $G$. For example the set of gates $G = \{\textsc{and, not}\}$ and $G = \{\textsc{nand, fanout}\}$ are universal. Classical computation is generally not reversible. A computation is reversible if for every output it is always possible to uniquely recover the corresponding input. Every classical computation, however, can be made reversible. To do this, one needs to replace every gate in a given circuit with its reversible version. This in turn can be done by constructing the reversible version of the gates in a universal set. 

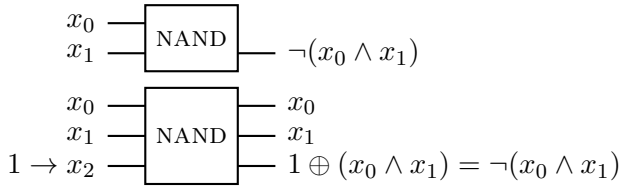
\begin{figure}[ht]
    \centering
    \begin{quantikz}[row sep = {0.4cm,between origins}]
        \lstick{$x_0$} & \gate[wires = 2]{\textsc{nand}} \\
        \lstick{$x_1$} & & \qw\rstick{$\neg (x_0 \wedge x_1)$} \\ [3mm]
        \lstick{$x_0$} & \gate[wires = 3]{\textsc{nand}} & \qw\rstick{$x_0$} \\
        \lstick{$x_1$} & & \qw\rstick{$x_1$} \\
        \lstick{$1 \rightarrow x_2$} & & \qw\rstick{$1 \oplus (x_0 \wedge x_1) = \neg (x_0 \wedge x_1)$}
    \end{quantikz}
    \caption{Irreversible (top) to reversible (bottom) \textsc{nand} gate, where $\neg$ denotes negation, $\oplus$ denotes Boolean operator \textsc{xor}, and $\wedge$ denotes Boolean operator \textsc{and}. The reversible gate is implemented using a particular setup of the Toffoli gate: the first two bits are control bits; the third one is a target bit, its input is always set to $1$.}
    \label{fig:rev-and}
\end{figure}

For example, the set $\{\textsc{nand, not}\}$ is universal. The gate $\textsc{not}$ is already reversible. The gate $\textsc{nand}$ can be made reversible by adding an additional input and two additional outputs. This is known as Toffoli gate~\cite{DBLP:conf/icalp/Toffoli80}, \cite[Section 1.4.1]{nielsen_chuang_2010}. The two new outputs keep a copy of the original inputs, see Figure \ref{fig:rev-and}.

Note that any reversible gate has the same number of inputs and output bits~\cite{DBLP:conf/icalp/Toffoli80}. This means that any reversible circuit has the same number of input and output bits. A bit is usually denoted by a two-dimensional column vector: $[1, 0]^T$ for the bit $0$ and $[0, 1]^T$ for the bit $1$. This notation is used so that one can describe the probabilistic model of computation as well. Encoding bits using vectors leads to the natural representation of gates as linear operators. More concretely, any gate can be represented as a matrix. For example, the \textsc{not} gate is the following matrix
\begin{equation}
    \label{equ:not}
    \textsc{not} = \begin{bmatrix} 0 & 1 \\ 1 & 0 \end{bmatrix},
\end{equation}
which flips $0$ to $1$ and vice versa:
\begin{equation*}
    \textsc{not} \begin{bmatrix} 0 \\ 1 \end{bmatrix} = \begin{bmatrix} 1 \\ 0 \end{bmatrix}, \quad  \textsc{not} \begin{bmatrix} 1 \\ 0 \end{bmatrix} = \begin{bmatrix} 0 \\ 1 \end{bmatrix}.
\end{equation*}
Therefore, any circuit can be represented as a matrix that operates on the input bits as a vector and produces the output bits as vector. This formulation of the circuit model of computation proves useful for the quantum model of computation as well.

\subsection{The quantum model of computation}

\subsubsection{The Dirac notation}

The mathematical formulation of quantum computing is based on that of quantum mechanics. The main mathematical objects used in quantum mechanics are complex Euclidean spaces. These are vector spaces over the complex numbers called Hilbert spaces. For quantum computing we only deal with finite dimensional Euclidean spaces. A complex Euclidean space of dimension $n$ is denoted by $\Complex^n$ where $\Complex$ is the set of complex numbers and the exponent $n$ means Cartesian product $n$ times. 

An element of $\Complex^n$ is called a vector and is usually denoted by an arrow over a letter, for example $\vec{x}$. In quantum mechanics, however, a vector is denoted by $\ket{x}$, and its adjoint is denoted by $\bra{x}$. This is called the Dirac notation, which we will also use in this paper. The standard basis of $\Complex^n$ is denoted by $\ket{0}, \ket{1}, \dots, \ket{n - 1}$, every vector is a unique linear combination of these vectors, i.e., $\ket{a} = a_0\ket{0} + \cdots + a_{n - 1}\ket{n - 1}$, where $a_i \in \Complex$ for all $i$. If $n$ is a power of two, say $n = 2^k$, then  basis vectors are written using binary strings of length $k$: $\ket{00 \dots 00}, \ket{00 \dots 01}, \dots, \ket{11 \dots 10}, \ket{11 \dots 11}$. This basis is usually referred to as the \textit{computational basis}. 

Other equivalent notations for a basis vector $\ket{b_1, \dots, b_n}$ in the computational basis are $\ket{b_1} \ket{b_2} \cdots \ket{b_n}$ or $\ket{b_1} \otimes \ket{b_2} \otimes \cdots \otimes \ket{b_n}$, where $\otimes$ is the tensor product operation. Tensor product is an operation that combines two spaces together. For example if $\{ \ket{a_i} \}_{1 \le i \le m}$ is a basis for the space $\Complex^m$ and $\{ \ket{b_j}_{1 \le j \le n} \}$ is a basis for the space $\Complex^n$ then $\{ \ket{a_i} \otimes \ket{b_j} \}_{1 \le i \le m, 1 \le j \le n}$ is a basis for the space $\Complex^m \otimes \Complex^n$.

\subsubsection{Qubits and quantum mechanics}

In the following, we briefly review the four postulates of quantum mechanics that form the conceptual foundations of quantum computing; see~\cite{kaye2007introduction,nielsen_chuang_2010} for additional details.

We start with the \textit{State Space Postulate}, which says that the state space of a 1-qubit quantum system is described by the set of unit vectors in $\Complex^2$. Therefore, in the computational basis, a qubit is described by a linear combination $\ket{\psi} = \alpha \ket{0} + \beta \ket{1}$ where $\abs{\alpha}^2 + \abs{\beta}^2 = 1$. We refer to $\ket{\psi}$ as a quantum state, in this case the state of a 1-qubit system. We also say that $\ket{\psi}$ is a \textit{superposition} of the states $\ket{0}$ and $\ket{1}$. The \textit{Composition of Systems Postulate} states that the combined system of two quantum systems is described by the tensor product of the corresponding state spaces. More precisely, if two quantum systems have state spaces $H_1$ and $H_2$ then the composite system has states space $H_1 \otimes H_2$. This means that, for example, a $2$-qubit system is described by the superposition of the basis states $\ket{00}, \ket{01}, \ket{10}, \ket{11}$. More generally, the state of an $n$-qubit system can be written as the superposition
\begin{equation}
    \label{equ:superpos}
    \ket{\psi} = \sum_{x \in \{ 0, 1 \}^n} \alpha_x \ket{x}
\end{equation}
of basis states, where $\alpha_x \in \Complex$ and $\sum_{x \in \{ 0, 1 \}^n} \abs{\alpha_x}^2 = 1$. The quantum state of a system can evolve, over time, to another quantum state. 

The \textit{Evolution Postulate} says that the state of a closed quantum systems evolves according to unitary operators. This means for any evolution of a system from a state $\ket{\psi_1}$ to a state $\ket{\psi_2}$ there exists a unitary operator $U$ such that $U \ket{\psi_1} = \ket{\psi_2}$. An operator is called unitary if $U^* \, U = I$, where $U^*$ is the adjoint of $U$, and $I$ is the identity operators. If we fix a basis for the state space of the quantum system, an operator is represented by a unique matrix. In that case, the adjoint of an operator is the conjugate-transpose of the corresponding matrix. As explained in the previous section, the evolution of classical systems can also be described by matrices. Therefore, loosely speaking, quantum operations can be thought of as a generalization of classical operations that act on continuous state spaces. A $1$-qubit operator is represented by a $2 \times 2$ matrix acting on the space $\Complex^2$. For example, for the computational basis, the unitary operator that takes the qubit $\alpha\ket{0} + \beta\ket{1}$ to $\alpha\ket{1} + \beta\ket{0}$ and vice versa, is the \textsc{not} operation in \eqref{equ:not}. Another example is given by a set of well-known $1$-qubit operators deemed the Pauli operators:
\begin{equation}
    \label{eq:pauli}
    \sigma_0 = \begin{bmatrix} 1 & 0 \\ 0 & 1 \end{bmatrix}, \; \sigma_1 = \begin{bmatrix} 0 & 1 \\ 1 & 0 \end{bmatrix}, \;  \sigma_2 = \begin{bmatrix} 0 & -i \\ i & 0 \end{bmatrix}, \; \sigma_3 = \begin{bmatrix} 1 & 0 \\ 0 & -1 \end{bmatrix}.
\end{equation}
Note that $\textsc{not} = \sigma_1$.

According to the \textit{Measurement Postulate}, quantum measurement on a system $A$ is described by a set of measurement operators which act on the state space of A. The state of the system and the probability of being in that state after the measurement depends on the measurement operators. A concrete example of quantum measurement is the von Neumann measurement with respect to the computational basis: given the state $\ket{\psi}$ in \eqref{equ:superpos}, performing a von Neumann measurement with respect to the basis $\{ \ket{x} \}$ outputs $y$ with probability $\abs{\alpha_y}^2$, and the state of the system after the measurement is $\ket{y}$. In general, performing a measure produces some classical information and leaves the system in a (possibly) new quantum state.

\subsubsection{The quantum circuit model}

\begin{figure}[ht]
    \centering
    \begin{quantikz}[row sep = {0.8cm,between origins}]
        \lstick{$q_0$} & \gate{G_1} & \qw & \qw & \gate[wires = 2]{G_5} & \qw \\
        \lstick{$q_1$} & \ctrl{-1} & \gate[wires = 2]{G_3} & \gate{G_4} & & \qw \\
        \lstick{$q_2$} & \gate{G_2} & & \meter{} & \qw & \qw
    \end{quantikz}
    \caption{A quantum circuit.}
    \label{fig:qtm-crct}
\end{figure}
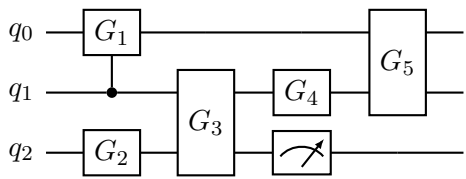

As explained in Section \ref{sec:cls-cmm-model}, any classical circuit can be efficiently converted to a reversible circuit. The reversible model of computation can naturally be generalized to a model of quantum computation. The quantum circuit model is similar to the reversible circuit model with bits and gates replaced by qubits and quantum gates: wires carry qubits to quantum gates, the gates perform quantum operations on the input qubits, and the resulting qubits are again carried by wires. Figure \ref{fig:qtm-crct} shows a schematic example of a quantum circuit.

Note that since quantum gates represent unitary operations, the number of input wires to a quantum circuit is the same as the number of output wires. This is not the case for classical circuits. As mentioned above, a $1$-qubit operator is a unitary operator that acts on the $2$-dimensional $\Complex^2$. Such an operator is called a $1$-qubit gate. When used as quantum gates, the Pauli operators $\sigma_0, \sigma_1, \sigma_2, \sigma_3$, defined in~\eqref{eq:pauli}, are denoted by $I, X, Y, Z$, respectively. The gate $X$ is quantum \textsc{not} gate. Just like classical gates, quantum gates can have \textit{control} inputs. For example, the control-\textsc{not} (\textsc{cnot}) gate acts on a $2$-qubits state as $\ket{a}\ket{b} \mapsto \ket{a}\ket{a \oplus b}$, where $\oplus$ is the \textsc{xor} operation. Here, the first qubit is the control qubit, based on the value of which a \textsc{not} gate is applied to the second qubit. Since \textsc{cnot} acts on $\Complex^4$, it is represented by a $4 \times 4$ matrix which, with respect to the computational basis, is
\[
\textsc{cnot} =
\begin{bmatrix}
    1 & 0 & 0 & 0 \\
    0 & 1 & 0 & 0 \\
    0 & 0 & 0 & 1 \\
    0 & 0 & 1 & 0
\end{bmatrix}.
\]
The measurement gate is usually depicted by a meter where the input is a quantum state and the output is classical information (and a post-measurement quantum state).
\begin{center}
    \begin{quantikz}
        \lstick{quantum state} & \meter{} & \qw\rstick{classical information}
    \end{quantikz}
\end{center}

The concept of universality can be generalized to quantum computation. A fundamental difference with the classical case is that the set of unitary operations is not discrete, hence a discrete set of quantum gates cannot be used to implement arbitrary unitary operations exactly. 

A set $G$ of quantum gates is called universal if any unitary operation can be approximated to arbitrary accuracy by quantum circuits involving only gates from $G$. It can be proved~\cite[Section 4.3]{kaye2007introduction} that the set $\{H, T\}$, where $H$ and $T$ are the Hadamard gate and the $\frac{\pi}{8}$-gate defined by
\[ H = \frac{1}{\sqrt{2}}\begin{bmatrix} 1 & 1 \\ 1 & -1 \end{bmatrix}, \quad T = \begin{bmatrix} 1 & 0 \\ 0 & e^{i\pi / 8} \end{bmatrix}, \]
is universal for $1$-qubit gates. It means that any $1$-qubit gate can be arbitrarily approximated by circuits involving only $H$ and $T$. If we add \textsc{cnot} to the above set, we obtain the universal set of gates $G = \{\textsc{cnot}, H, T\}$ for quantum computation~\cite[Section 4.3]{kaye2007introduction}.

\section{Debugging Quantum Programs}\label{sec:debug_quantum}

A quantum program executed on a modern gate-based QC leverages a register of qubits for performing quantum operations and a register of classic bits for recording the measurements of qubits' states and conditionally applying quantum operators~\cite{cross2017open}. Thus, a typical QC program mixes traditional instructions (to alter the state of bits and apply conditional statements) and quantum instructions (to alter the state of qubits and to measure qubit value).

As mentioned above, a general quantum program consists of blocks of code each containing classical and quantum instructions. 
Quantum operations can be divided into two kinds: unitary and non-unitary. Unitary operations are 
reversible and preserve the norm of the operands. Non-unitary operations are not reversible and have 
probabilistic implementations. 

The classical parts of a quantum program can be debugged using traditional methods. The quantum parts, 
however, can not be treated in the same way because of the properties of a QC~--- such as superposition, 
entanglement, and no-cloning~--- which are governed by the laws of quantum mechanics. The purpose of 
debugging a program is to present the user with human readable, i.e., classical, information about 
the runtime state of the system. Extracting classical information from a quantum state is done using 
measurement which is non-unitary and results in collapse of the state, and hence 
an unintended behavior of the program. We shall describe, in the following, different scenarios in a 
QC to which classical debugging techniques cannot be applied, and discuss some potential solutions.

\subsection{Superposition}\label{sec:superposition}

\begin{figure}[ht]
	\centering
	\begin{subfigure}[b]{0.25\columnwidth}
		\centering
		\includegraphics[width = \textwidth]{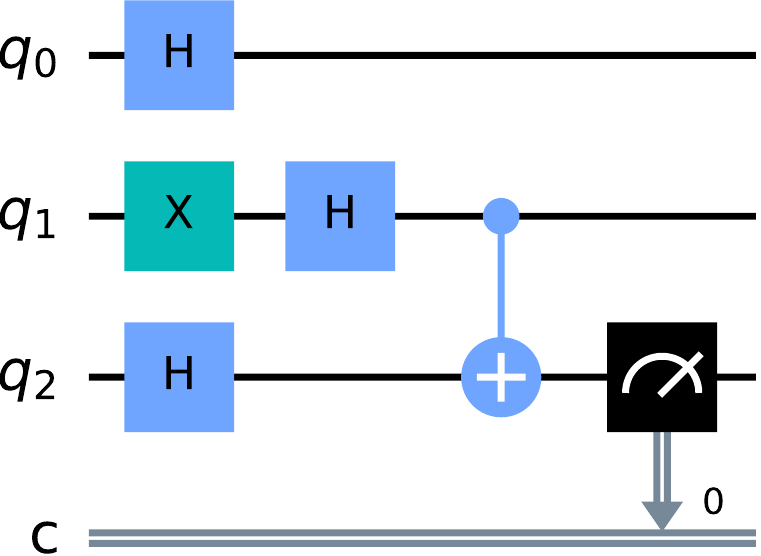}
		\caption{Circuit}
		\label{fig:spp-circ}
	\end{subfigure}
	\hspace*{1cm}
	\begin{subfigure}[b]{0.25\columnwidth}
		\centering
		\begin{minted}[fontsize = \footnotesize, numbersep = 2mm, linenos = true, autogobble]{cpp}
            OPENQASM 2.0;
            include "qelib1.inc";
            
            qreg q[3];
            creg c[1];
            
            x q[1];
            h q[0];
            h q[1];
            h q[2];
            cx q[1],q[2];
            measure q[2] -> c[0];
        \end{minted}
        \vspace{-4mm}
		\caption{Assembly code}
		\label{fig:spp-code}
	\end{subfigure}
	\caption{Example of measuring a superposition. In OpenQASM, \textsc{not} is denoted by $x$, Hadamard by $h$, and \textsc{cnot} by $cx$. }
	\label{fig:spp}
\end{figure}

Let $\ket{\psi}$ be the state of an $n$-qubit register. Then we can uniquely write $\ket{\psi}$ using a superpostion in the computational basis as in \eqref{equ:superpos}. By 
the measurement postulate of quantum mechanics, measuring the state $\ket{\psi}$ in the 
computational basis results in an outcome $x \in \{ 0, 1 \}^n$ with probability $\abs{\alpha_x}^2$, 
and the state of the system after the measurement is $\ket{x}$. For example, consider the initial state
$\ket{010}$ and perform the following steps: first apply a Hadamard transform to each qubit (creating superposition), then
a controlled-not \textsc{cnot} to qubits 2 and 3, and finally measure qubit 3. If the measured qubit is $0$ 
(which happens with probability $1/2$), then the state collapses to $\frac{1}{2}(\ket{00} - \ket{01} 
+ \ket{10} - \ket{11})$. An implementation of this example in OpenQASM 2.0 is shown in Figure \ref{fig:spp}.

A natural feature of a debugger for quantum programs would be to check if the state 
of a variable is in superposition. There are two possible scenarios: when the input state is unknown (e.g., when it is generated as an output of another quantum program) and when the input state is known. Let us elaborate on each of these cases.

\subsubsection{Unknown input state.}\label{sec:unknown-inp}
If the input to the program is an unknown state $\ket{\psi}$, then there is no known general 
algorithm that can efficiently decide if $\ket{\psi}$ is in a superposition. 
Not much can be done here in terms of a general method for debugging, 
different approaches should be considered for different problems.

For example, in the hidden subgroup problem \cite[Chapter 7]{kaye2007introduction}, if the group is abelian, then it can be efficiently 
decided if the coset state of a subgroup is in superposition. For non-abelian groups, however, the same problem is often hard. For example, 
the best known algorithm for the following problem has subexponential runtime \cite{kuperberg2005subexponential}: let $N$ be a positive 
integer, and let $\Z_N$ be the group of integers mod $N$. For a random unknown $x \in \Z_N$ and fixed unknown 
$d \in \Z_N$, decide whether a given state is of the form $\ket{b}\ket{x}$ or $\frac{1}{\sqrt{2}}(\ket{0}\ket{x} + \ket{1}\ket{x + d})$, where $b \in \{0, 1\}$.

\subsubsection{Known input state.}\label{sec:known-inp}
\begin{figure*}[ht]
	\centering
	\begin{subfigure}[b]{0.45\columnwidth}
		\centering
		\includegraphics[width = \textwidth]{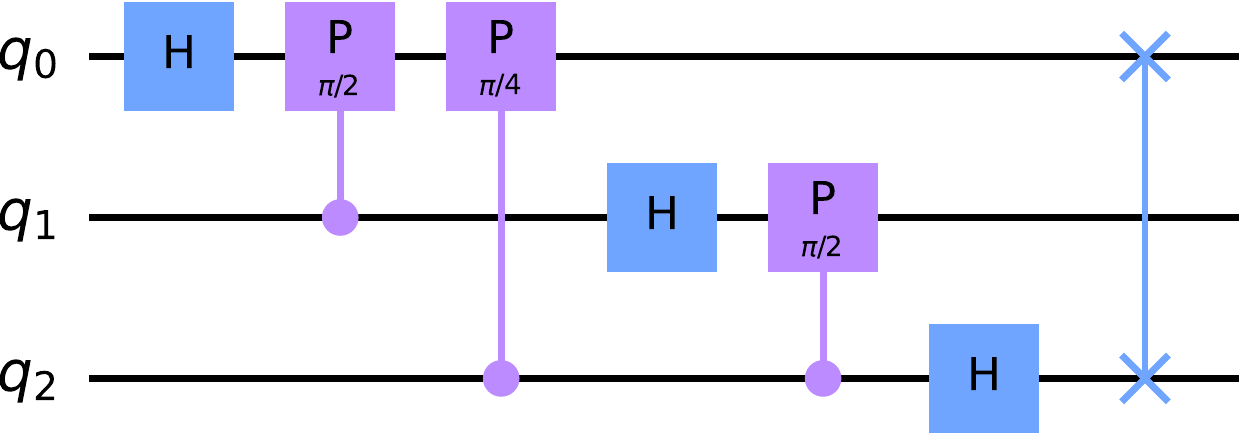}
		\caption{Circuit}
		\label{fig:qft-circ}
	\end{subfigure}
	\hspace*{10mm}
	\begin{subfigure}[b]{0.25\columnwidth}
		\centering
		\begin{minted}[fontsize = \footnotesize, numbersep = 2mm, linenos = true, autogobble]{cpp}
            OPENQASM 2.0;
            include "qelib1.inc";
            
            qreg q[3];
            
            h q[0];
            cp(pi/2) q[1],q[0];
            cp(pi/4) q[2],q[0];
            h q[1];
            cp(pi/2) q[2],q[1];
            h q[2];
            swap q[0],q[2];
        \end{minted}
        \vspace{-4mm}
		\caption{Assembly code}
		\label{fig:qft-code}
	\end{subfigure}
	\caption{Quantum Fourier Transform over $\Z_8$. In OpenQASM, $P_{\pi / 2}$ and $P_{\pi / 4}$ phase gates are denoted by $cp(\cdot)$; \textsc{swap} gate by $swap$. }
	\label{fig:qft}
\end{figure*}

If a state is the result of applying a known unitary operation to a known initial state, i.e., 
$\ket{\psi} = U \ket{\psi_0}$ where $\ket{\psi_0}$ and $U$ are both known, then $\ket{\psi}$ can be regenerated by the 
debugger. For example, consider the state 
\[ \ket{\psi} = \frac{1}{\sqrt{2^n}} \sum_{x \in \{ 0, 1 \}^n} (-1)^{h(x)} \ket{x}, \]
where $h(x)$ is the Hamming weight of $x$, i.e., the number of non-zero bits in $x$. Then $\ket{\psi}$ can be generated by applying the Hadamard transform $U = H^{\otimes n}$ to the $n$-qubit register $\ket{11 \dots 1}$. This is an example of the Quantum Fourier Transform (QFT) over the group $\Z_2^{\oplus n}$. QFT can be implemented efficiently over the group $\Z_N$ where $N$ is an integer, although in this case the implementation is more involved. A toy implementation of QFT for $N = 8$ is shown in Figure \ref{fig:qft}. 

In the cases as above, there are various methods (depending on 
the problem) to characterize the state $\ket{\psi}$. Often, one relies on \textit{quantum state 
tomography}, which is the process of reconstructing a quantum state through a series of measurements \cite{d2003quantum, cramer2010efficient}.

\subsection{Entanglement}

In a QC, a set of memory cells or registers is said to be in an entangled state if it is impossible 
to classically specify the correlations among them. More precisely, let $X_1, \dots, X_n$ be the 
state spaces of a set of quantum systems that represent $n$ registers. The state space of the 
composite of these systems, that represents an array, is given by the tensor product $X = X_1 
\otimes \cdots \otimes X_n$. A state $\ket{\psi} \in X$ that can be written in the form 
$\ket{\psi} = \ket{\psi_1} \otimes \cdots \otimes \ket{\psi_n}$, where $\ket{\psi_j} \in 
X_j$ for $j = 1, \dots, n$, is called separable. A state that is not separable is called entangled. 
When debugging a program that operates on an entangled state, the following problems can be 
considered.

\subsubsection{Checking for separability.}\label{sec:separability}
Given a state $\ket{\psi} \in X$, deciding whether $\ket{\psi}$ is separable is an NP-hard 
problem \cite{gharibian2010strong, gurvits2003classical}. This is called the \textit{separability 
problem} in quantum information theory, see \cite[Chapter 6]{watrous2018theory} for details. There are a variety of methods (see~\cite{leinaas2006geometrical,guhne2009entanglement}) for separability/entanglement detection 
that can be implemented in practice, specially for lower dimensions. For example, if the debugger can generate 
several copies of $\ket{\psi}$, then one way to detect the nonlinear properties of 
$\ket{\psi}$ is via direct measurement. For the sake of brevity, we do not provide technical details here; see \cite{leinaas2006geometrical} for a numerical method 
for examining separability and \cite{guhne2009entanglement} for other interesting methods and their 
implementations.

\subsubsection{Extracting classical information.}\label{sec:classical}
Measuring a subsystem of a larger composite system that is in an entangle state will likely alter 
other subsystems. In debugging terms, if a set of variables are in an entangled state, a debugger will not be able to present any classical information about a subset of those variables to the user without disturbing the state of the whole set. 

For example, consider the entangled state $\frac{1}{\sqrt{2}}(\ket{00} + \ket{11})$ of two qubits.
Measuring any of the two qubits alters the result of the subsequent measurement on the other qubit.
More precisely, if the first qubit is measured, then state collapses to $\ket{00}$ or $\ket{11}$ with probability $|1/\sqrt{2}|^2 =
1/2$; the outcome of measuring the second qubit is always $0$ if the resulting state is $\ket{00}$, and it is always $1$ if the resulting state is $\ket{11}$. Such a state is called maximally entangled. 

A composite system, however, often has subsystems that are not entangled with any other subsystem. In 
this case, we can measure that subsystem without disturbing the whole state. For example, in the following state of a 
3-qubit register
\begin{equation}
\label{equ:sep-subsys}
	\frac{1}{2} (\ket{000} - \ket{001} + \ket{110} - \ket{111}),
\end{equation}
the last qubit is not entangled with the first two while the first and the second qubits are 
entangled, see \eqref{equ:gen-clas}. The algorithms for separability detection (discussed in Section~\ref{sec:separability})  could be used to identify separable subsystems. 

\begin{figure}[ht]
	\centering
	\begin{subfigure}[b]{0.25\columnwidth}
		\centering
		\includegraphics[width = \textwidth]{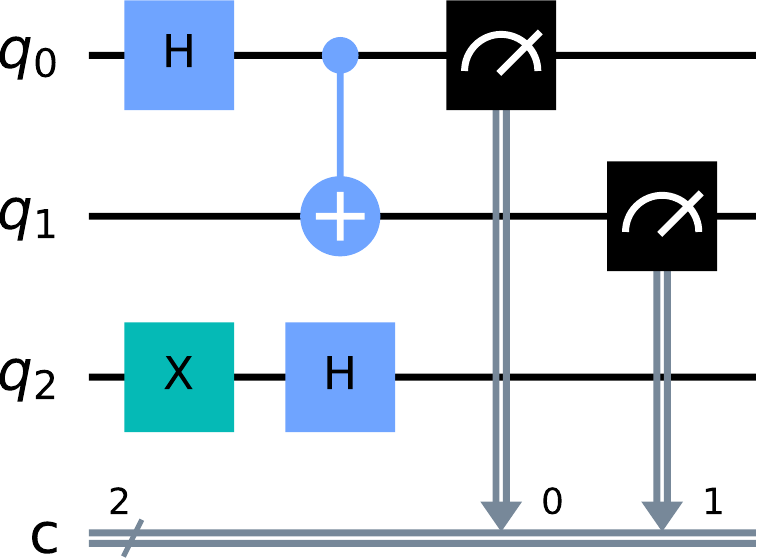}
		\caption{Circuit}
		\label{fig:entg-circ}
	\end{subfigure}
	\hspace*{1cm}
	\begin{subfigure}[b]{0.25\columnwidth}
		\centering
		\begin{minted}[fontsize = \footnotesize, numbersep = 2mm, linenos = true, autogobble]{cpp}
            OPENQASM 2.0;
            include "qelib1.inc";
            
            qreg q[3];
            creg c[2];
            
            x q[2];
            h q[0];
            cx q[0], q[1];
            h q[2];
            measure q[0] -> c[0];
            measure q[1] -> c[1];
        \end{minted}
        \vspace{-4mm}
		\caption{Assembly code}
		\label{fig:entg-code}
	\end{subfigure}
	\caption{A circuit for generating state \eqref{equ:sep-subsys}.}
	\label{fig:entg-sep}
\end{figure}

Things would be much simpler if the debugger could somehow estimate a given state with a state that 
is generated by applying some operation to a basis state (i.e., classical information). For example, 
the state in \eqref{equ:sep-subsys} can be generated as
\begin{equation}
    \label{equ:gen-clas}
    \begin{aligned}
        & (\textsc{cnot} \otimes H)(H \otimes I_4) \ket{001} \\
        & = \frac{1}{\sqrt{2}}(\ket{00} + \ket{11}) \otimes \frac{1}{\sqrt{2}}(\ket{0} - \ket{1}),
    \end{aligned}
\end{equation}
where $I_4$ is the $4 \times 4$ identity gate. Therefore, the state in \eqref{equ:sep-subsys} can be described by the debugger using the classical information 
$\ket{001}$ and the names of the above operators. An implementation of the sequence of operations in \eqref{equ:gen-clas} is 
shown in Figure \ref{fig:entg-sep}.

\subsection{No-cloning}\label{sec:no-cloning}

The most general method of obtaining information about a variable without disturbing its state is 
to make a copy of the variable and work on the copy. In the classical setting, this is often 
straightforward. In the quantum setting, however, the situation is much more complicated. In fact, 
it is impossible to make a copy of a given general unknown quantum state. More precisely, given an 
unknown state $\ket{\psi}$ and an arbitrary state $\ket{\phi}$, it can be shown~\cite[Theorem 10.4.1]{kaye2007introduction} that 
there is no unitary operator $U$ that can perform the following:
\[ \ket{\psi} \otimes \ket{\phi} \overset{U}{\longmapsto} \ket{\psi} \otimes 
\ket{\psi}. \] 
In many practical scenarios, however, a debugger will only need to make an \textit{approximate copy} 
of a state; a state that is `close enough' to the given state but provides useful debugging 
information. For example, for a state $\ket{\psi}$ that encodes a probability distribution \cite{grover2002creating}, such as
the Gaussian distribution, an approximate clone would provide valuable information about the distribution.
The possibility of approximate cloning was first discussed in \cite{buvzek1996quantum}. 
Much research has been done on different cloning methods each optimizing particular aspects of a 
cloner that are desired for different situations, see \cite{scarani2005quantum} for a survey. 

\begin{figure}[ht]
	\centering
	\begin{subfigure}[b]{0.25\columnwidth}
		\centering
		\includegraphics[width = \textwidth]{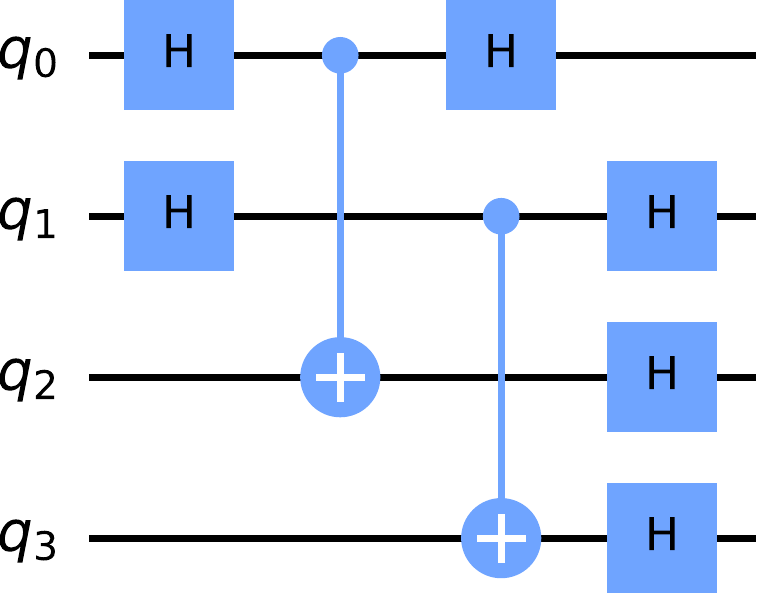}
		\caption{Circuit}
		\label{fig:clone-orth-circ}
	\end{subfigure}
	\hspace{1cm}
	\begin{subfigure}[b]{0.25\columnwidth}
		\centering
		\begin{minted}[fontsize = \footnotesize, numbersep = 2mm, linenos = true, autogobble]{cpp}
            OPENQASM 2.0;
            include "qelib1.inc";
        
            qreg q[4];
        
            h q[0];
            h q[1];
            cx q[0],q[2];
            h q[0];
            cx q[1],q[3];
            h q[1];
            h q[2];
            h q[3];
        \end{minted}
        \vspace{-4mm}
		\caption{Assembly code}
		\label{fig:clone-orth-code}
	\end{subfigure}
	\caption{A circuit for cloning $2$-qubit states in the set $S$. Here, the state of the first two qubits $q_0, q_1$ is copied into the last two qubits $q_2, q_3$.}
	\label{fig:clone-orth}
\end{figure}

Also, it is possible to perform exact cloning if the given state belongs to a set of known mutually orthogonal states. Recall that two states $\ket{\psi_1}$, $\ket{\psi_2}$ are orthogonal if $\braket{\psi_1}{\psi_2} = 0$. For example, consider the set of $n$-qubit states $S = \{ \ket{\psi_j} \}$ where
\[ \ket{\psi_j} = \frac{1}{\sqrt{2^n}} \sum_{x \in \{ 0, 1 \}^n} (-1)^{\lrang{j, x}} \ket{x}, \]
for $j \in \{0, 1\}^n$. The set $S$ is exponentially large, but for any given state $\ket{\psi_j} \in S$ where $j$ is unknown, we can efficiently make a copy of $\ket{\psi_j}$. An example of such a cloning procedure, for $n = 2$, is shown in Figure \ref{fig:clone-orth}.

\subsection{Discussion}
In Sections~\ref{sec:superposition}--\ref{sec:no-cloning}, we discussed various issues preventing the application of the classic debugging techniques and identified some potential solutions. 

As discussed in~\cite{miranskyy2019testing}, if the input size and the amount of required qubits is small, we can run a quantum program in a simulator (running on a CC). However, the increase of the input size and the qubit register length may force us to run the program on a QC. 

If we can generate multiple approximate copies of the state \cite{buvzek1996quantum}, then we can produce an empirical distribution of the qubit state and compare it against the expected distribution, to detect problems in the code. The generation of the multiple approximate copies can be readily implemented for moderate inputs sizes using universal cloning methods~\cite{werner1998optimal, buvzek1998universal, fan2001quantum}.  More efficient cloning can be achieved using state-dependent (i.e. non-universal) cloning methods~\cite{niu1999two, scarani2005quantum}. This would address issues related to superposition with known input state (discussed in Section~\ref{sec:known-inp}), extraction of classical information (discussed in Section~\ref{sec:classical}), and no-cloning (discussed in Section~\ref{sec:no-cloning}). A compiler can automatically generate the code for the approximate copying (akin to compilers for CC that can instrument the code to add debugging information), translating higher-level language into quantum assembly~\cite{huang2019statistical}. The same principle of multiple approximate copies (aggregated using statistics) can be used to generate runtime assertions~\cite{li2014debugging,zhou2019quantum,zhou2019quantum_extended,DBLP:journals/pacmpl/LiZYDY020}.

For the case of unknown input states, discussed in Section~\ref{sec:unknown-inp}, no general solution exists and will require a programmer to make decisions on a case-by-case basis.

Finally, separability checking, discussed in Section~\ref{sec:separability}, demands the implementation of numerical methods that will require changes to the QC and, hopefully, will be implemented in the future. 

\section{Conclusions}\label{sec:conclusions}

QC field is rapidly evolving, and the SE community should start bringing SE practices into the QC world.
In this paper, we focus on analyzing testing and debugging tactics, highlighting classic ones that are readily applicable and showing that new ones have to be created.
We believe that this work would be of interest to practitioners, creating quantum programs, as well as researchers, developing the next generations of tooling for QC.

\section*{Acknowledgments}
We express profuse thanks to the anonymous reviewers of this paper and of~\cite{miranskyy2019testing} and~\cite{DBLP:conf/icse/Miranskyy0D20}. We are also grateful to the ICSE 2020 committee for rewarding~\cite{DBLP:conf/icse/Miranskyy0D20} with the New Ideas and Emerging Results Distinguished Paper Award.

\bibliography{references}

\end{document}